\documentclass[%
preprint,
showpacs,
amsmath,amssymb,
aps,
prb,
]{revtex4-1}

\usepackage[super]{nth}
\usepackage{textcomp}
\usepackage{graphicx}
\usepackage{epstopdf}

\begin{document}

\preprint{}

\title{Magnetotransport Measurements of the Surface States of Samarium Hexaboride using Corbino Structures}

\author{S. Wolgast}
\email{swolgast@umich.edu}
\thanks{These authors contributed equally to this work}
\author{Y. S. Eo}
\email{eohyung@umich.edu}
\thanks{These authors contributed equally to this work}
\author{T. \"{O}zt\"urk}
\altaffiliation[Also at ]{Sel\c{c}uk University, Dept.~of Physics, Konya, 42075, Turkey.}
\author{G. Li}
\author{Z. Xiang}
\altaffiliation[Also at ]{Hefei National Laboratory for Physical Science at Microscale and Department of Physics, University of Science and Technology of China, Hefei Anhui 230026, China.}
\author{C. Tinsman}
\author{T. Asaba}
\author{B. Lawson}
\author{F. Yu}
\author{J. W. Allen}
\author{K. Sun}
\author{L. Li}
\author{\c{C}. Kurdak}
\affiliation{University of Michigan, Dept.~of Physics, Ann Arbor, Michigan 48109-1040, USA}

\author{D.-J. Kim}
\author{Z. Fisk}
\affiliation{University of California at Irvine, Dept.~of Physics and Astronomy, Irvine, California 92697, USA}

\date{September 4, 2015}

\begin{abstract}
The recent conjecture of a topologically-protected surface state in SmB$_6$ and the verification of robust surface conduction below 4 K have prompted a large effort to understand the surface states. Conventional Hall transport measurements allow current to flow on all surfaces of a topological insulator, so such measurements are influenced by contributions from multiple surfaces of varying transport character. Instead, we study magnetotransport of SmB$_6$ using a Corbino geometry, which can directly measure the conductivity of a single, independent surface. Both (011) and (001) crystal surfaces show a strong negative magnetoresistance at all magnetic field angles measured. The (011) surface has a carrier mobility of $122\text{ cm}^2/\text{V}\cdot\text{sec}$ with a carrier density of $2.5\times10^{13} \text{ cm}^{-2}$, which are significantly smaller than indicated by Hall transport studies. This mobility value can explain a failure so far to observe Shubnikov-de Haas oscillations.  Analysis of the angle-dependence of conductivity on the (011) surface suggests a combination of a field-dependent enhancement of the carrier density and a suppression of Kondo scattering from native oxide layer magnetic moments as the likely origin of the negative magnetoresistance. Our results also reveal a hysteretic behavior whose magnitude depends on the magnetic field sweep rate and temperature. Although this feature becomes smaller when the field sweep is slower, does not disappear or saturate during our slowest sweep-rate measurements, which is much slower than a typical magnetotransport trace. These observations cannot be explained by quantum interference corrections such as weak anti-localization, but are more likely due to an extrinsic magnetic effect such as the magnetocaloric effect or glassy ordering.
\end{abstract}


\maketitle

\section{\label{sec:Intro}Introduction}

Samarium hexaboride (SmB$_6$) has captured renewed interest due to recent theoretical predictions\cite{Dzero,Takimoto} suggesting that it is a strong 3D topological insulator (TI), and also due to subsequent experimental verifications of a conducting surface state consistent with TI surface states predicted for the material\cite{Wolgast,Kim}, as well as evidence from tunneling spectroscopy\cite{ZhangJP}. As a result, there have been a large number of theoretical calculations\cite{FengLu,Alexandrov,MYe} and experimental works providing strong evidence that the surface conduction has a TI surface state contribution. Hybridization gap and metallic surface formation have been studied by a wide range of spectroscopic measurements, including angle-resolved photoemission spectroscopy (ARPES)\cite{Miyazaki,NXu,JJiang,Neupane,Denlinger1,Denlinger2,Zhu,Xu_SpinARPES}, point-contact spectroscopy\cite{ZhangJP}, and scanning tunneling spectroscopy\cite{Yee, Rossler, Ruan} experiments. The reported hybridization gap values are slightly different, but are roughly in agreement at $\sim20$ meV. De Haas-van Alphen (dHvA) oscillations have also been observed on distinct 2D crystallographic surfaces\cite{GLi}, although the size of the Fermi pockets reported are mostly not in agreement with ARPES results. As is expected for a TI, the surface states can be suppressed through bulk magnetic impurity doping\cite{KimMag,WangAPS2014}. Even though the topological nature of the system has recently been called into question\cite{Rader}, we nonetheless adopt the TI framework for our analysis, based on the wealth of experimental evidence and the compelling general theoretical basis for this scenario.

SmB$_6$ stands out from the known 3D semiconductor TIs because its topological band structure arises from strong electron correlation effects. Furthermore, because SmB$_6$ has a fully insulating bulk below $\sim4\text{ K}$, the electrical properties of the surface states can be probed directly and easily by transport measurements; this is not possible for the known 3D semiconductor TIs due to polluting conductivity from the bulk. Normally for a two-dimensional electron gas (2DEG), Shubnikov-de Haas oscillations (SdH) can be used to extract the carrier density ($n_{\text{2D}}$) and mobility ($\mu_{\text{2D}}$). However, so far there is no convincing evidence of SdH oscillations up to 45 T\cite{GLi}, suggesting that the surface has a low mobility (on the order of $100\text{ cm}^{2}/\text{V}\cdot\text{sec}$ or lower). However, there are reports of weak anti-localization\cite{SThomasWAL,NakajimaJP} (WAL), as expected for a TI. 

Although a Hall bar structure is typically used to characterize magnetotransport of both 2D and 3D conductive states, 3D TIs pose particular difficulties for this conventional geometry. All surfaces of the Hall bar contribute to the total conduction, including any edges or corners that are not perpendicular to the magnetic field, and may vary in surface condition due to preparation procedures such as polishing. For example, this can lead to an effective ``edge channel'' that would short the quantum Hall insulator state of the surfaces perpendicular to the field. Another complication arises if the surface states exhibit ambipolar conduction, as is indicated in calculations by Lu \textit{et al}.\cite{FengLu}. The Hall coefficient is sensitive to charge sign, and in a multi-channel scenario with both electron and hole conduction, the contributions of one to the Hall coefficient can compensate the other. Our own Hall bar measurements on SmB$_6$ indicated\cite{Wolgast} carrier densities that were unphysically large for a 2D system, perhaps because any or all of these complications reduced the measured value of the Hall coefficient. Unfortunately, these complications also now make the large volume of past detailed low-temperature transport work in SmB$_6$ (which assumed the low-temperature resistivity plateau to be a bulk effect) very difficult to interpret, especially since details about crystal size and geometry are usually not reported.

In this paper, we avoid these particular difficulties by fabricating Corbino disks on single surfaces of SmB$_6$. This geometry is not sensitive to the sign of the charge(s), and is sensitive only to the surface on which it is fabricated. The longitudinal conductivity, $\sigma_{xx}$, of the surface can be directly obtained from the 2-terminal resistance and the geometry of the disk. There is a geometrical diminution of $\sigma_{xx}$ under a perpendicular magnetic field because the current begins to circulate, lengthening the path over which an average charge carrier must travel through the system. The conductivity of a single-carrier system is then given by
\begin{equation}\label{eq:corb}\sigma_{xx}(B)=\frac{ne\mu}{1+\mu^2B^2_{\perp}},\end{equation}
where $n$ is the carrier density of the surface, $\mu$ is the carrier mobility, and $B_{\perp}$ is the perpendicular component of the magnetic field. This dependence on the magnetic field allows us to obtain values for $\mu$ and $n$.

In this work, we measure $\sigma_{xx}(B)$ using a Corbino geometry on the (001) and (011) surfaces of SmB$_6$. The magnetoresistance (MR) and angle dependence we observe at $0.3\text{ K}$ at high magnetic fields are consistent with a picture where Kondo scattering off magnetic impurities immediately adjacent to the surface dominates the transport behavior of the surface states at low fields, and where the high-field MR is due to increases in $n$, accompanied by a small decrease in $\mu$ via short-range disorder scattering. Meanwhile, our resistivity data at low magnetic fields exhibit dynamic hysteretic behaviors which become stronger at faster magnetic field sweep rates. At a fixed sweep rate, the feature caused by this hysteresis resembles WAL. However, the strong sweep-rate dependence suggests that this feature is not caused by a quantum interference effect, but rather by a magnetic effect. We attribute this dynamically slow hysteresis to an extrinsic effect such as the magnetocaloric effect or a glassy magnetic ordering of the magnetic impurities near the surface. The native samarium sesquioxide (Sm$_{2}$O$_{3}$) formed on the surface after exposure to ambient air is a likely source for such surface magnetic impurities, which influence the low-field transport behavior of SmB$_6$.

\section{\label{sec:Methods}Experimental Methods}

Single-crystal SmB$_{6}$ samples were grown by the Al flux method. Typical pieces had $1-2$ mm $\times$ $600-1000$ \textmu m surfaces, and were thinned in the (001) or (011) crystallographic direction to $300-500$ \textmu m thicknesses by manual polishing with coarse SiC grit or by automated lapping using Al$_{2}$O$_{3}$ slurry. We polished the surface of interest on each piece with SiC abrasive pads (grit size P4000) or 0.3 \textmu m slurry. We lithographically patterned the Corbino disks with an inner diameter of 300 \textmu m and an outer diameter of 500 \textmu m. We ashed the surfaces with oxygen plasma and evaporated 50/1500 {\AA} Ti/Au contacts, followed by lift-off of the active region. We attached wires to the contacts using Au or Al wirebonding, reinforcing the contacts with silver paint for better adhesion where needed. For most of our samples, two wires for each source and drain were bonded so that the resistance of the wires could be neglected when performing four-terminal measurements. One of our samples with a complete Corbino disk with contacts is shown in the inset of Figure~\ref{fig:HighFieldMR}. Contact resistances were Ohmic both at 300 K and at 4 K.

We performed alternating-current resistance measurements at high magnetic fields using standard lock-in techniques in multiple magnet systems at the National High Magnetic Field Laboratory (NHMFL). Angle-dependent resistance measurements were performed in the NHMFL 35 Tesla system using constant currents of 2 \textmu A and 5 \textmu A for the (011) and (001) surfaces, respectively. Measurements at lower fields were taken in a $^3$He cryostat with an 8 T superconducting magnet and a $^3$He/$^4$He dilution refrigerator with a 14 T superconducting magnet, both using a bipolar magnetic power supply, whose current polarity switching occurs at $B\neq0\text{ T}$. These resistance measurements were also taken using standard lock-in techniques, and in some cases, a pre-amplifier and a bridge circuit were also used to achieve clearer signals. The time constant that determines the low-pass filter bandwidth of the lock-in amplifier was set short enough ($\tau=1\text{ sec}$) so that even at our fastest magnetic field sweep rates ($32\text{ mT/sec}$), the associated time delay is not significant. The excitation current ($I=10^{-7}$ to $10^{-6}\text{ A}$) was sufficiently small that the measured resistance did not depend on the current or frequency.

\section{\label{sec:HResults}Measurements of Magnetoresistance at Large Magnetic Fields}

\begin{figure}[b]
\includegraphics{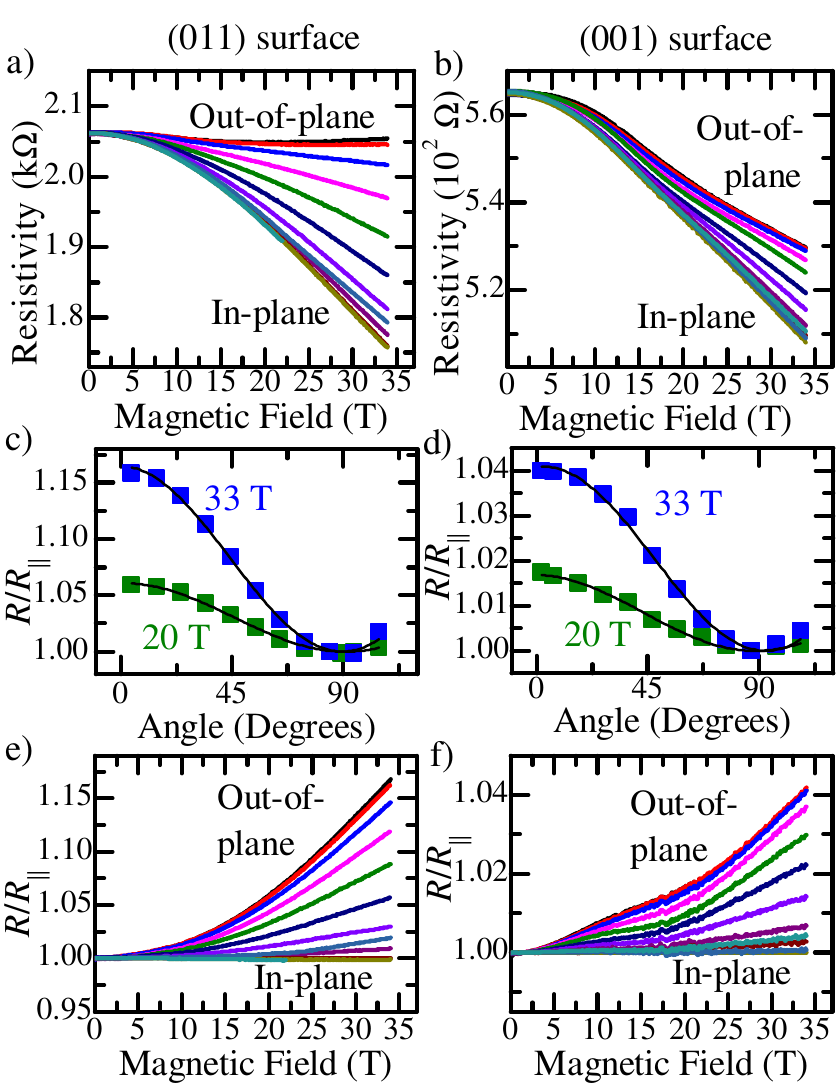}
\caption{(Color online) Magnetoresistance traces at several angles for the (a) (011) surface and the (b) (001) surface. Angle-dependence of the ratio (points) of the resistance with out-of-plane field ($R$) to the resistance with in-plane field ($R_{\parallel}$) for the (c) (011) surface and the (d) (001) surface at representative magnetic fields, along with $\cos^2\theta$ fits (lines). The ratio $R/R_{\parallel}$ is plotted vs. magnetic field for the (e) (011) and (f) (001) surfaces.}
\label{fig:RRC}
\end{figure}

Figure~\ref{fig:RRC} shows MR traces obtained at 0.3 K for multiple field angles, measured with respect to the surface normal, in the NHMFL 35 Tesla system for the (011) and (001) surfaces of SmB$_6$, respectively. The most apparent feature for both surfaces is the strong negative MR at all measured angles. We also note that we do not observe Shubnikov-de Haas (SdH) oscillations for either surface up to 45 T, which is perhaps surprising in light of the observation of dHvA oscillations at lower field values by our collaborators\cite{GLi}.

One of the most striking features of the traces is their angle-dependence, which is primarily a result of the perpendicular field-dependence of $\sigma_{xx}$ arising from the Corbino geometry and included in the denominator of Equation~\ref{eq:corb}. Taking the ratio of traces for in-plane magnetic field and magnetic field with arbitrary angle~$\theta$ with respect to the surface normal eliminates $n$ and gives
\begin{equation}\label{eq:ratio}\frac{\sigma(B_{\parallel})}{\sigma(B)}=1+\mu^2B^2\cos^2(\theta),\end{equation}
from which we can directly obtain $\mu$. The ratio associated with each surface is plotted for different magnetic fields as a function of angle in Figure~\ref{fig:RRC}~(c) and (d), and for each angle as a function of magnetic field in Figure~\ref{fig:RRC}~(e) and (f). Both sets of ratios exhibit an apparent $\cos^2\theta$ dependence, which is the expectation for a surface conduction in the Corbino geometry (Eq.~\ref{eq:ratio}), and the (011) ratios also approximately exhibit the expected $B^2$ dependence. Simple quadratic fits of the (011) field-dependent curves in Figure~\ref{fig:RRC}~(e) yield a carrier mobility of $122 \text{ cm}^2/\text{V}\cdot\text{sec}$ and a carrier density of $2.5\times10^{13}\text{ cm}^{-2}$. Both of these values are much lower than previously reported for Hall bar transport measurements\cite{Wolgast,Kim}, which may suffer from the problems discussed earlier. However, they are both more consistent with values from ARPES measurements\cite{Neupane,JJiang,Denlinger2} and other Corbino disk experiments\cite{SyersJP}, and the carrier density value is physically plausible. Such a low mobility suggests that SdH oscillations will not be detectable below $1/\mu=81\text{ T}$, which explains why we do not observe them. Meanwhile, the (001) ratios do not exhibit a simple $B^2$ dependence, most likely due to the presence of multiple carrier channels which may have different MRs. A two-carrier formulation in which the channels have similar conductivities but very different carrier mobilities will yield a total $\sigma(B_\parallel)/\sigma(B)$ with a shape similar to the data ratios in Figure~\ref{fig:RRC}~(f), but it will not quite fit the data without additional MR-related contributions to each channel. However, our data does not sufficiently constrain the parameters of such a multiple-carrier fit with MR. Thus, in the rest of this section, we will limit our focus to the (011) surface, except where noted.

The MR, which is not explicitly included in Equations~\ref{eq:corb} or~\ref{eq:ratio}, is due to $B$-dependence of $n$, $\mu$, or both. A more detailed analysis allows us to investigate the relative contributions of $n(B)$ and $\mu(B)$ to the MR. The coupling between the orbital motion of 2D surface electrons and the external magnetic field is expected to show a $\cos^2\theta$ dependence similar to that of Eqs.~\ref{eq:corb} and~\ref{eq:ratio}, and would not affect $\sigma(B_{\parallel})$. Meanwhile, other mechanisms (e.g. contributions from the Zeeman splitting) are expected to be independent (or only weakly dependent) on $\theta$. Because most of the $\theta$-dependence in the data comes from the Corbino geometry, and because $\sigma(B_{\parallel})$ exhibits large MR, we proceed with the assumption that $n(B)$ and $\mu(B)$ are independent of the field angle $\theta$. (We note that a small $\theta$-dependent contribution is expected to arise from the weakening of TI backscattering suppression due to the magnetic field's influence on the helical spin dispersion\cite{Ozturk}, but we calculate that this effect is negligible at the field values measured here.) We plot the carrier densities and mobilities obtained from $\cos^2\theta$ fits at constant $B$ (e.g., Figure~\ref{fig:RRC}~(c) and (d)) as a function of magnetic field (symbols in Figure~\ref{fig:AngleFits011}). The high quality of the fits (see Appendix~\ref{apx:001Cond}) at large $B$ supports the assumption that $n(B)$ and $\mu(B)$ are sufficiently independent of $\theta$ such that $n(B)$ and $\mu(B)$ can then be obtained with good precision. However, the fits (and the analytical form of Eqs.~\ref{eq:corb} and \ref{eq:ratio}, solved for $n(B)$ and $\mu(B)$) are divergently sensitive to noise near $B=0$, so this method does not work well at low field values, which is evident in the uncertainty of the values in Figure~\ref{fig:AngleFits011}.

\begin{figure}
\includegraphics{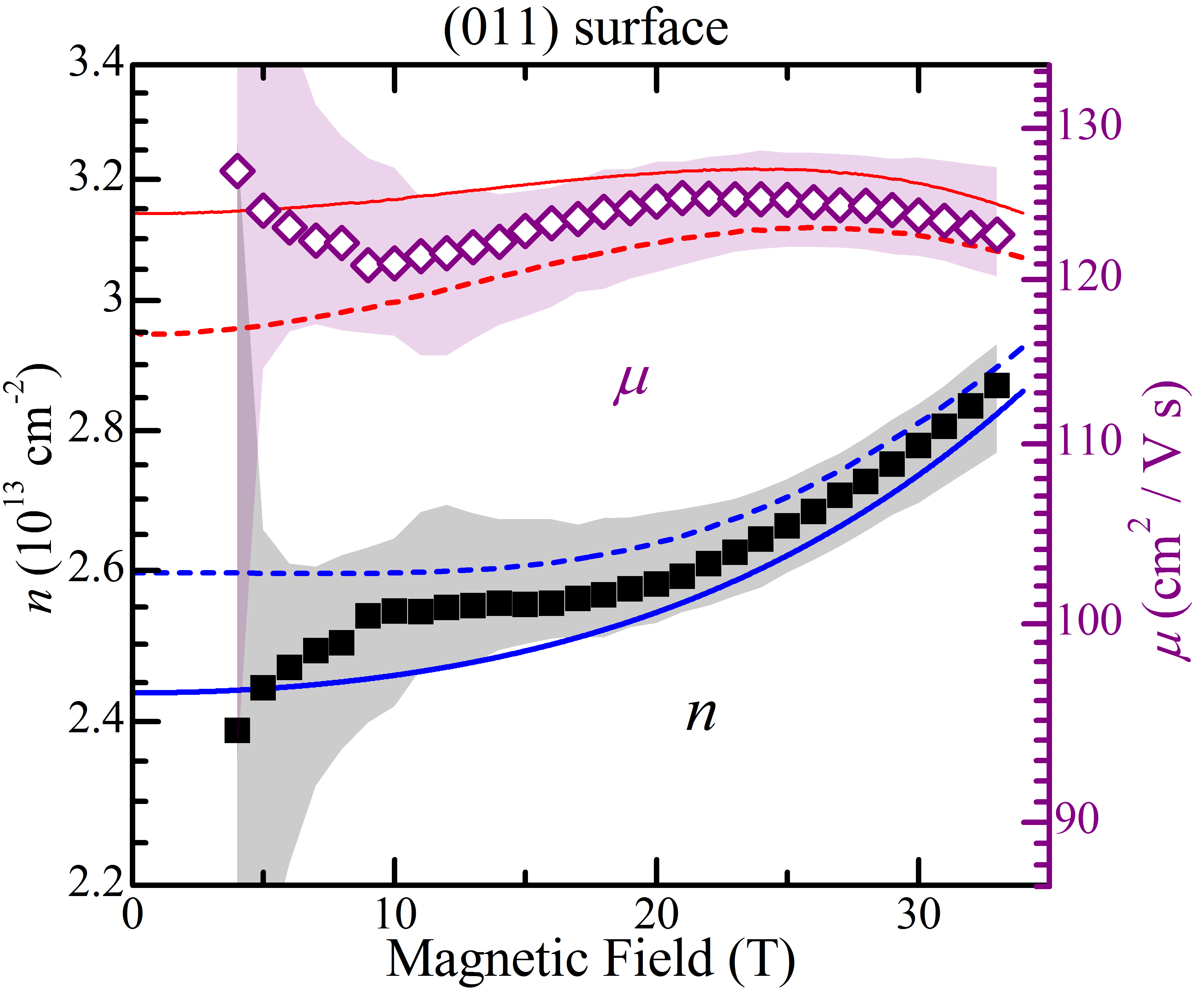}
\caption{(Color online) (011) surface carrier density (filled squares) and mobility (open diamonds) obtained from angle-dependent fits of the data. Shaded areas represent uncertainty in the parameters of the angle-dependent fits. Best-fit curves for polynomial $n(B)$ (blue) and corresponding $\mu(B)$ (red) using the $\theta=25^{\circ}$ and $\theta=85^{\circ}$ data (solid lines), and using the $\theta=5^{\circ}$ and $\theta=85^{\circ}$ data (dotted lines). The vertical log scale allows direct comparison of the relative magnitudes of changes in $n$ and $\mu$.}
\label{fig:AngleFits011}
\end{figure}

To overcome this problem at low $B$, we assume that $n(B)$ can be approximated using an even polynomial in $B$. By treating the polynomial coefficients as fitting parameters, we can determine a best fit for $n(B)$ and $\mu(B)$, constrained by two $\sigma(B)$ traces at different $\theta$ of our choosing. Solid (dotted) lines in Figure~\ref{fig:AngleFits011} show the best fit for a \nth{6}-order polynomial $n(B)$ using the $\theta=85^{\circ}$ trace and the $\theta=25^{\circ}$ ($\theta=5^{\circ}$) trace, along with the corresponding $\mu(B)$. Fits at other angles change the relative magnitude of $n(B)$ and $\mu(B)$ by $<10\%$, suggesting some small angle-dependence of $n(B)$ and $\mu(B)$ that is not captured in our two-parameter model, but the qualitative dependence on $B$ remains the same. Both the $\theta$-dependent fits and the $B$-dependent fits suggest that changes in carrier density are primarily responsible for the MR of the (011) surface; i.e., the MR is a result of large changes in the carrier density accompanied by small changes in the carrier mobility.

For the (001) surface, a na\"{i}ve application of single-carrier $\cos^2\theta$ fits above $25\text{ T}$ yields a constant mobility of $61\text{ cm}^2/\text{V}\cdot\text{sec}$ and an increasing carrier density around $2\times10^{14}\text{ cm}^{-2}$. If such fits are taken at face value, they suggest that the (001) surface's MR is also dominated by changes in carrier density. However, below $\sim25\text{ T}$, the fit residuals start becoming much larger. Meanwhile, a polynomial best-fit of $n(B)$ fails to reproduce the $B$-dependence of the data, giving credence to the notion that the analysis is complicated by the presence of multiple carrier channels with different MRs or another unknown $\theta$-dependent effect (see Appendix~\ref{apx:001Cond}).

Multiple channels giving rise to visible MR features at distinguishable magnetic field values is an indication that the channels likely have carrier densities and mobilities that differ by orders of magnitude, but have resistivities of the same order. In fact, this is a reasonable expectation in a system that exhibits both large and small Fermi pockets, as has been observed on the (001) surface of SmB$_6$ both by ARPES and dHvA measurements. In one possible scenario, the large pocket, which is centered about the X point and has a large carrier density, may suffer from short-range disorder scattering and have a comparatively small mobility, as discussed later. Meanwhile, the small pocket, which is centered around the $\Gamma$ point and has a smaller carrier density, may be dominated by long-range impurity scattering, which allows a much higher mobility.

\section{\label{sec:LResults}Hysteretic Magnetotransport at Small Magnetic Fields}

\begin{figure}
\includegraphics{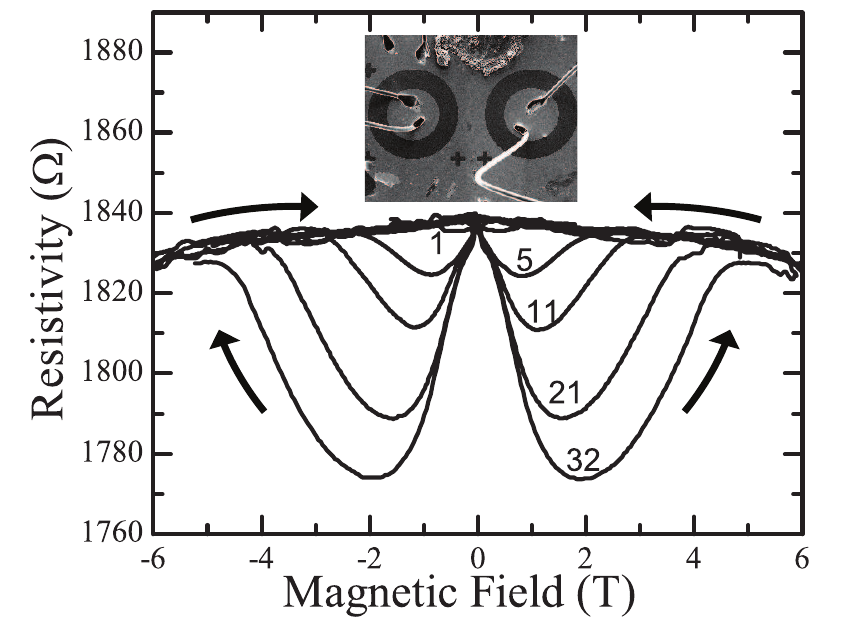}
\caption{Response of resistivity of the Corbino disk samples to the magnetic field at different sweep rates below 6 T at $0.3\text{ K}$. The numbers shown close to each curve are the magnetic field sweep-rate magnitude in units of mT/sec. The inset on top of trace shows an example of a Corbino disk sample image prepared on a polished SmB$_{6}$ surface.}
\label{fig:HighFieldMR}
\end{figure}

We now focus specifically on low magnetic fields, for which the response of the resistivity shows slow dynamical hysteretic behaviors. Specifically, the resistivity is dependent on the history of the magnetic field and its sweep rate. For a systematic study, we start from a large magnetic field value ($-B_{\text{max}}$ to $+B_{\text{max}}$) and measure resistivity, sweeping in both directions at different field sweep rates ($dB/dt$). Figure~\ref{fig:HighFieldMR} shows typical resistivity traces of one of our Corbino disk samples at different sweep rates. This dynamical hysteretic behavior was observed in most of our samples. Following the arrows in this figure, while sweeping the magnetic field from $-6$ T ($-B_{\text{max}}$) until $0$ T, the resistivity does not show any strong features. However, continuing from $0$ to $+6$ T ($+B_{\text{max}}$), a noticeable dip occurs. The resistivity first starts to decrease and reaches to some minimum value. Then, the resistivity starts to return to its path as the magnetic field is further increased. When the sweep direction is reversed and the field is swept from $+6$ T ($+B_{\text{max}}$) to $0\text{ T}$, this dip does not appear. Continuing from 0 to $-6$ T ($-B_{\text{max}}$), the strong dip appears again. As a result, the two strong dips appear symmetrically on each polarity of the magnetic field. By increasing the magnetic field sweep rate, the magnitude of these dips becomes larger. Typically, these dips appear at magnetic fields smaller than $\pm5\text{ T}$. We note that hysteresis has also been reported by other workers\cite{NakajimaJP} at similar magnetic fields, but there are significant qualitative differences between those results and ours.

\begin{figure}
\includegraphics{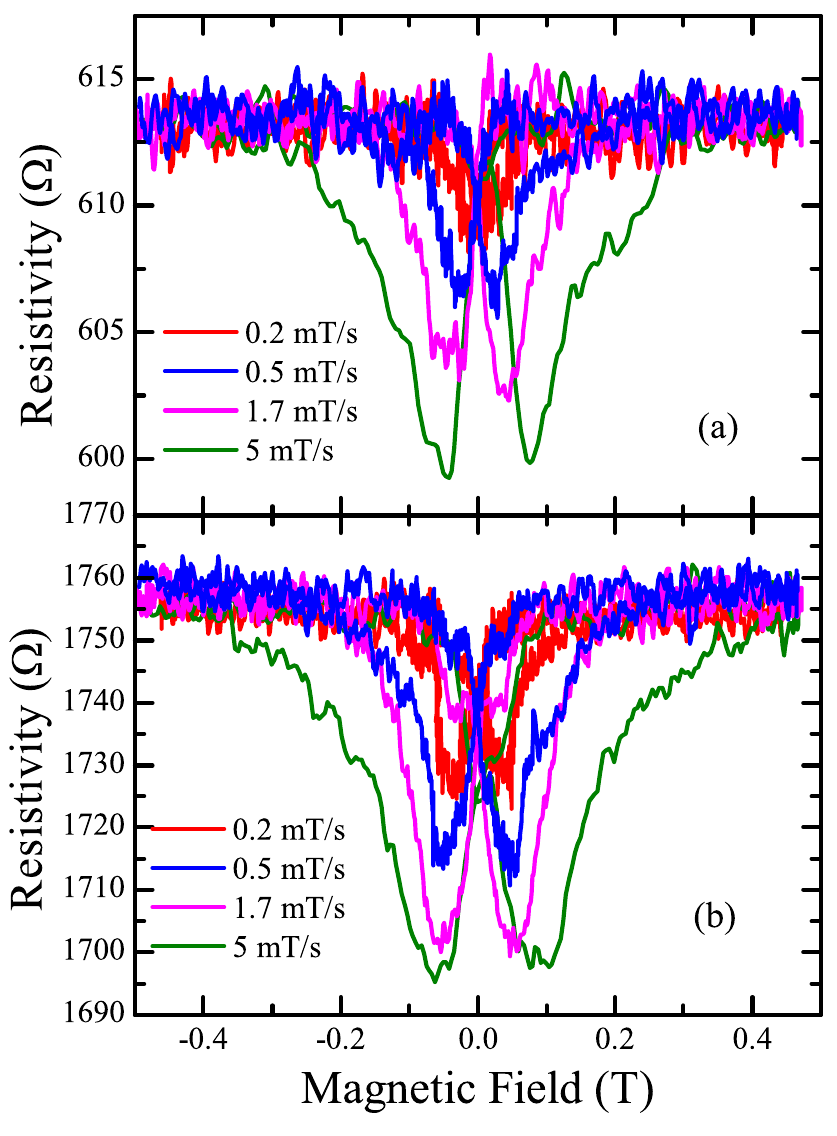}
\caption{(Color) Response of resistivity of the Corbino disk samples to the magnetic field at different sweep rates below 1 T at 80 mK. (a) (001) sample (b) (011) sample.}
\label{fig:LowFieldMR}
\end{figure}

\begin{figure}
\includegraphics{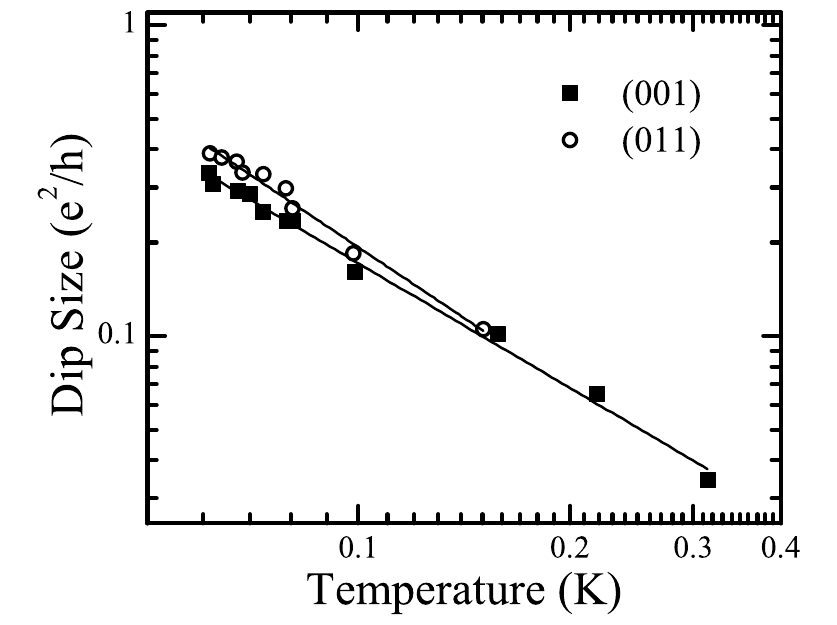}
\caption{Magnitude of the dips (in conductivity) as a function of temperature at magnet field sweep rate $0.167$ mT/s.}
\label{fig:DipTemp}
\end{figure}

\begin{figure}
\includegraphics{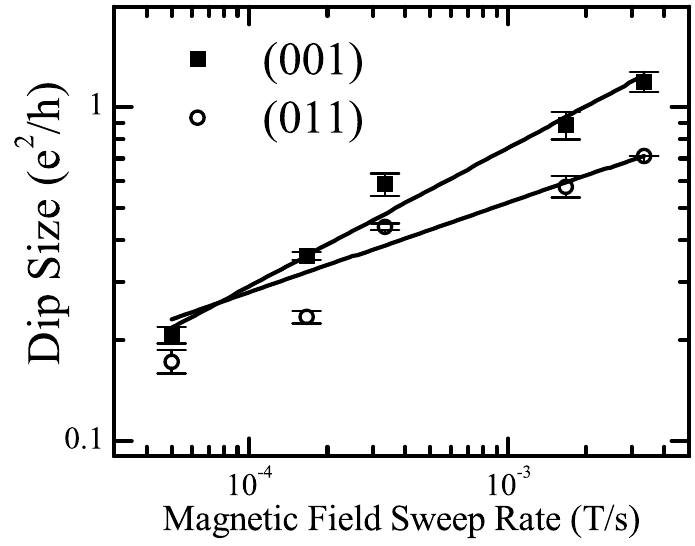}
\caption{Magnitude of the dips (in conductivity) as a function of magnet field sweep rate at 80 mK.}
\label{fig:DipSweep}
\end{figure}

\begin{figure}
\includegraphics{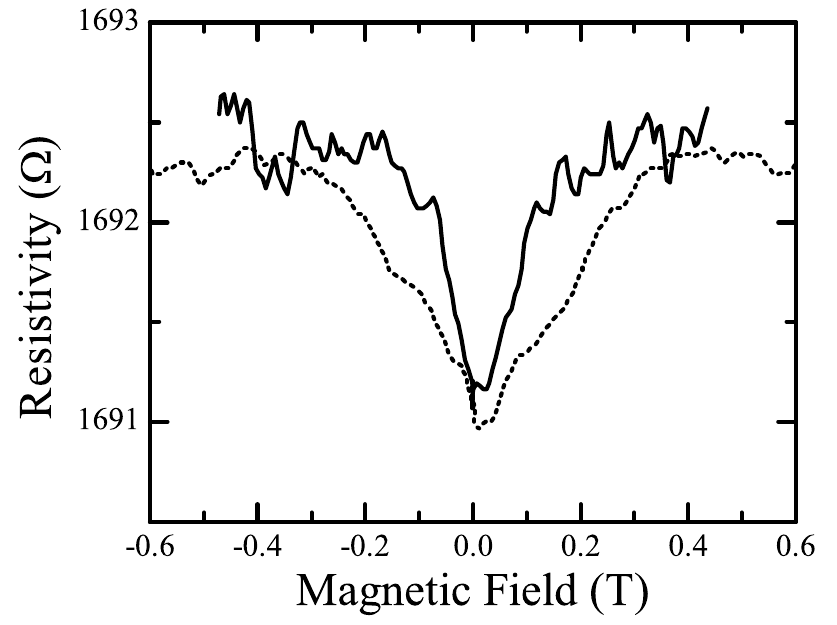}
\caption{Response of resistivity of the Corbino disk samples to the magnetic field comparing at different angles of magnetic field at $0.3$ K. Solid curve is the magnetic field perpendicular to the transport surface. Dotted curve is magnetic field parallel to the transport surface. The minimal points are shifted to $B$ = 0 T for direct comparison.}
\label{fig:Angle}
\end{figure}

We also observe an additional hysteretic feature in all of our samples at lower magnetic field ranges (within $\pm$ 1 T). As shown in Figure~\ref{fig:LowFieldMR}, these features were systematically studied on two samples as described above at lower temperatures (down to 60 mK) and extremely slow magnetic field sweep rates (down to 0.2 mT/s). The features have two symmetric dips similar to those in Figure~\ref{fig:HighFieldMR}. The hysteretic features are smaller, and the positions of the minima appear at a lower field range, but the qualitative magnetic field response remains the same. Previously, WAL has been reported\cite{SThomasWAL, NakajimaJP} within this range. However, the sweep-rate dependent dynamic dips that we observe in Figure~\ref{fig:LowFieldMR} are not caused by WAL. For the WAL case, the magnetic field only breaks the phase of the electrons traveling a closed loop by scattering off static impurities, and this phase does not depend on $dB/dt$. If one were to fix on a particular magnetic field sweep rate, the data do show some similarities to WAL. As shown in Figure~\ref{fig:DipTemp}, when converting the magnitude of the dips ($\Delta R$) to change in conductivity ($\Delta\sigma$), the sizes are on the order of typical WAL peak magnitudes ($\sim0.1 e^{2}/h$). Also similar to WAL, $\Delta\sigma$ increases as the temperature is lowered. However, the magnetic field response must be static for WAL. Although the magnitude of the dips decreases at slower sweep rates, we did not observe any sign of the dip magnitude saturating (becoming non-dynamic). The magnitude of the dips as a function of magnetic field sweep rate for both samples is shown in Figure~\ref{fig:DipSweep}. Even at the slowest measurements ($dB/dt=5\times10^{-5}\text{ T}$), which takes more than 5 hours to sweep 1 T, the magnitude of the dips continues to shrink. In addition to this measurement, we took angle-dependent magnetic field measurements that also indicate that this feature is not WAL. WAL can only be observed as a function of the perpendicular magnetic field component\cite{HikamiLarkinNagoaka}. As shown in Figure~\ref{fig:Angle}, however, the dips also appear in parallel (in-plane) magnetic fields, and this dip widens very slowly compared to what we expect from a typical WAL feature as the field is rotated from the perpendicular to the parallel direction\cite{HikamiLarkinNagoaka}. 

Since the hysteretic features we observed are not WAL, we must assume that if WAL exists, it is buried under the hysteretic dips. For this to happen, since the WAL features are static, they must be smaller than the smallest hysteretic dip size that we observed ($\sim0.2e^2/h$ in Figure~\ref{fig:DipSweep}). We can estimate the expected magnitude of WAL, which is calculated to be $\delta\sigma=(\alpha/2\pi)(e^2/h)\ln(\tau_{\phi}/\tau_{p})$\cite{Andersonlocalization, HikamiLarkinNagoaka}, where $\tau_{\phi}$ is the phase coherence time, $\tau_{p}$ is the momentum relaxation time, and $\alpha$ is the number of (identical) conduction channels. At the low temperature range we measured ($1\text{ K -- }60\text{ mK}$), $\tau_{\phi}$ can be theoretically estimated\cite{Altshuler,Fukuyama}, ranging on the order of $0.1\text{ -- }1.0\text{ ns}$. Calculating $\tau_{p}$ requires the unknown effective mass, $m^\ast$, in addition to the mobility we extracted from our high field measurements $(\tau_{p}=m^\ast\mu/e)$. If we use the effective mass from the recent measurements of dHvA and ARPES measurements\cite{GLi,Miyazaki,NXu,JJiang,Neupane,Denlinger1,Denlinger2,Zhu,Xu_SpinARPES}, where the effective mass is an order of magnitude smaller than the electron mass $(m^\ast\sim0.1m_{e})$, the WAL feature magnitude must be larger than the dip sizes of our hysteretic peaks $(\delta\sigma\sim0.7\text{ -- }0.89\times\text{number of conduction channels})$, which is inconsistent with our results. The effective mass must be much larger than the electron mass $(m^\ast\gg m_{e})$ for the dip size to be on the order of $0.1e^{2}/h$ or smaller. In the following section, we instead discuss a more plausible scenario which can also explain the absence of WAL as partly due to the presence of magnetic impurities.

\section{\label{sec:Discuss}Discussion}

\subsection{\label{sec:NegMR}Negative Magnetoresistance}

We now address the possible physical origins of the negative MR. Past measurements\cite{Cooley99} at 4 K have also observed strong negative MR. These researchers, assuming they were measuring fully bulk properties, attributed the negative MR to closure of the bulk gap $\Delta$ and an increase in $n_{\text{bulk}}$. Indeed, 4 K is very near the reported\cite{Wolgast,Kim} crossover temperature between surface-dominated and bulk-dominated conduction for similar flux-grown crystals. However, our own data is taken well below this transition temperature in a regime where the bulk is electrically dead, and the conduction we measure is purely due to the surface states. In this regime, the carrier density of the bulk bands is not related to the surface conduction, and a change in activated bulk transport with gap reduction is unable to explain the negative MR we observe. (Although we cannot rule out a bulk gap closure mechanism, our measurements are taken at a temperature at least one full order of magnitude below the crossover temperature. Pollution from bulk conduction at even 0.01\% does not arise in an activated transport model until the band gap is only 13\% of its zero-field value, which can happen no lower than $\sim80\text{ T}$\cite{Cooley99}. Additionally, we note that transport measurements taken at 4 K, which is near the crossover temperature, may be sensitive to any MR coming from the surface states.) It is, however, possible that a change in the bulk structure could have some effect on the surface states at the Fermi level (especially a change in the Dirac point relative to the Fermi energy), causing a change in the surface state carrier density. Because the fits of our data indicate that $n(B)$ is the dominant source of the negative MR, it seems reasonable to attribute the negative MR to such a bulk-driven ($\theta$-independent) picture. However, our collaborator's dHvA measurements\cite{GLi} suggest that the carrier density does not change significantly up to 45 T for any $\theta$. This disagreement, along with the large variations among reported values for $n$, $\mu$, and $k_{\text{F}}$ from ARPES studies\cite{Neupane,JJiang,Denlinger2}, remains to be resolved. We note that it is difficult to compare values from the transport studies and the ARPES studies, since the ARPES is performed in high vacuum, while the transport samples are exposed to ambient air.

\subsection{\label{sec:Kondo}Kondo Scattering}

\begin{figure}
\includegraphics{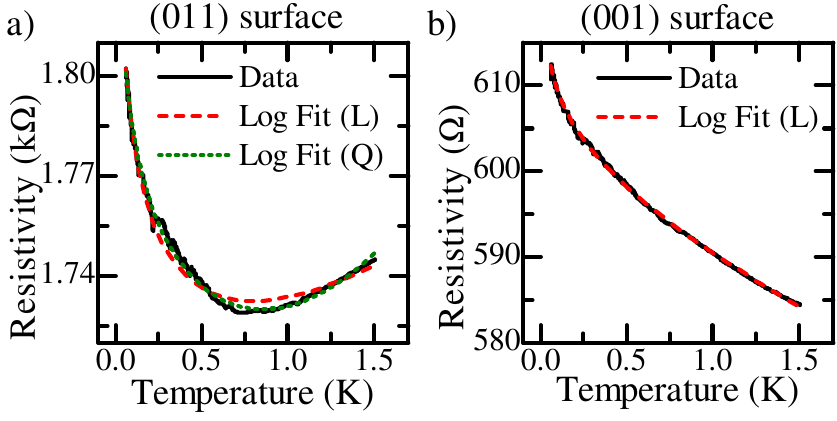}
\caption{(Color online) Resistivity vs. temperature for the (a) (011) surface and the (b) (001) surface. The solid black line is data, the long-dashed red lines are logarithmic fits on a linear temperature background, and the short-dashed green line is a logarithmic fit on a quadratic temperature background.}
\label{fig:Temp}
\end{figure}

If we take the $B$-dependence of $n$ as a given, we can investigate weaker features of the MR that are apparent in $\mu(B)$. Motivated by the observation of magnetic hysteresis at low fields (Section~\ref{sec:LResults}), we investigate magnetic impurity scattering as a likely contribution to the negative MR. We measured the Corbino resistances as a function of temperature with $B=0$ (Figure~\ref{fig:Temp}). On both surfaces, as the temperature is reduced, we observe a logarithmic increase of the resistance, the coefficient of which is far from the quantum conductance $e^2/h$. This, taken together with the low-field increase in $\mu(B)$, suggests a TI surface Kondo scattering mechanism\cite{Andrei,Xin}. We expect that there are significant magnetic impurities on the SmB$_6$ surface, based on recent X-ray magnetic circular dichroism and X-ray absorption spectroscopy spectra which show that Sm$^{3+}$ with a net magnetic moment is dominant on the surface\cite{Phelan}. In addition, hard X-ray photoelectron spectroscopy (HAXPES) shows a weak oxygen signal of a polished and then etched SmB$_{6}$ sample\cite{HaoTjeng}. These results imply that Sm$_{2}$O$_{3}$ oxide is formed when the surface of SmB$_{6}$ is exposed to air at ambient conditions\cite{HaoTjeng}. The native Sm$_{2}$O$_{3}$ formed on the SmB$_{6}$ surface is expected to be disordered. We therefore assess Kondo scattering from disordered Sm$^{3+}$ moments as a possible mechanism to explain both the temperature dependence and the low-field enhancement of $\mu(B)$.

\begin{figure}[h]
\includegraphics{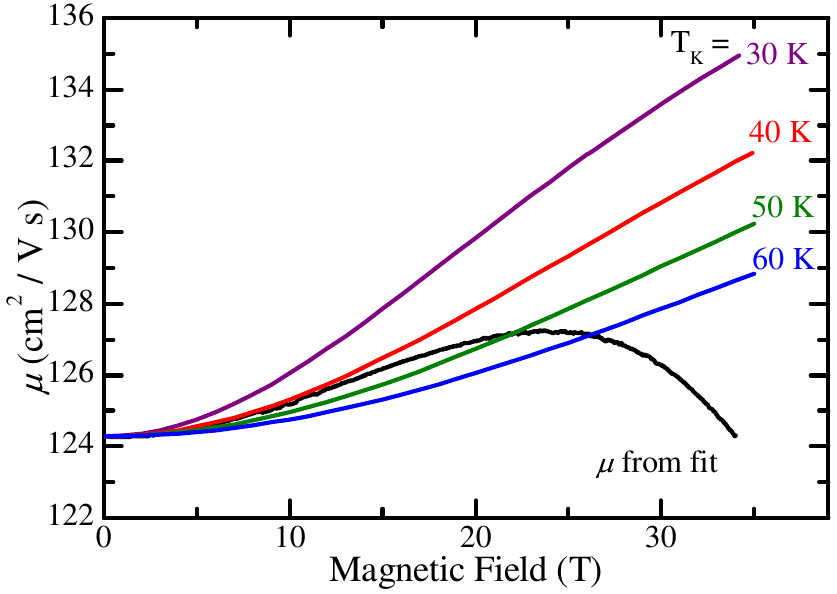}
\caption{(Color online) Fitted mobility, alongside several mobility projections of the Kondo effect for various $T_\text{K}$.}
\label{fig:Kondo}
\end{figure}

The logarithmic $T$-dependence of the electron scattering rate we observe can be fit using the following formula\cite{Xin}, developed for a 3D TI system with dilute magnetic impurities:
\begin{equation}\frac{1}{\bar{\tau}}\propto3+J\rho\ln\frac{T}{T_\text{K}},\end{equation}
where $J$ is the coupling constant, $\rho$ is the density of states at the Fermi energy, $T$ is the temperature, and $T_\text{K}$ is the Kondo temperature calculated using the renormalization group approach. Here, $\bar{\tau}$ represents the scattering-angle-averaged scattering time, since the spin-momentum locking of the TI surface states causes $\tau$ to depend on the scattering angle. For the SmB$_6$ surface, $J$ plausibly would arise from hybridization between the surface states and the paramagnetic Sm$_2$O$_3$ $f$-states. The Kondo scattering produces negative MR according to the formula\cite{Andrei,Xin},
\begin{equation}\frac{1}{\mu}=\frac{1}{\mu_{\text{d}}}+\frac{1}{\mu_{\text{M}}}\left(3+J\rho\ln\frac{T}{T_\text{K}}\right)\cos^2\left(\frac{\pi}{2}M(B)\right),\end{equation}
where $\mu_{\text{M}}$ is the coefficient of the contribution from Kondo scattering, $\mu_{\text{d}}$ is the mobility from disorder scattering alone, and $M(B)$, whose relationship to $B$ can be exactly calculated at low temperatures\cite{Andrei}, is the normalized magnetization of the impurities. (We note that this formulation may be quantitatively different from a 2D Kondo scattering description that might be more appropriate for the case of surface magnetic impurities. However, the qualitative behavior will be the same, which is sufficient here because we take the magnitudes of $\mu$, $\mu_{\text{d}}$, and $\mu_{\text{M}}$ as fitting parameters in the subsequent analysis.) Motivated by the experimental signatures of Kondo scattering, we apply this theory to SmB$_6$, even though the surface magnetic moments from the Sm$_2$O$_3$ might not be in the dilute limit. At zero magnetic field ($B$, $M=0$), the logarithmic fits shown in Figure~\ref{fig:Temp}, which include a linear background resistance of unknown origin, allow us to experimentally determine $\mu_{\text{d}}$ and $\mu_{\text{M}}$. The dependence on magnetic field ($B$, $M\neq0$), which arises from the suppression of spin-flip scattering due to Zeeman splitting, can then be predicted as a function of $T$ and $T_\text{K}$. Using the values from our logarithmic fits and for our value of $n(B=0)$, we plot in Figure~\ref{fig:Kondo} computed values for $\mu(B)$ for several different Kondo temperatures, alongside our fit of our experimental $\mu(B)$ for comparison. The low-field increase in $\mu(B)$ fits quite well with Kondo scattering for $T_\text{K}=40\text{ K}$; however, this is only an estimate, since other effects such as short-range scattering (discussed below) can also influence $\mu(B)$. We note that if we were to ignore the evidence for $B$-dependent $n$, and instead attribute the MR solely to changes in $\mu(B)$, such a na\"{\i}ve fit would yield a much larger negative MR than can be explained by Kondo scattering alone (see Appendix~\ref{apx:002Kondo}); the theoretical prediction and the experimental curve agree only by combining the Kondo effect together with the $B$-dependence of the carrier density.

We now consider a qualitative picture in which an oxide layer with dense magnetic impurities can still lead to Kondo scattering behavior. Such a Kondo lattice is formed by conduction carriers from the SmB$_6$ surface interacting with a disordered dense array of localized moments from the Sm$_2$O$_3$. If we first consider an ordered Kondo lattice, as the temperature is lowered from high temperatures, the resistivity rises logarithmically as the magnetic ordering becomes quenched by Kondo cloud formation, where the spin scattering between the localized $f$-electron and the $d$-conduction electron inside the cloud increases. As the temperature is lowered further, the resistivity drops since the effect of coherence between the lattice sites (Bloch's theorem) dominates, and the magnetic moment becomes quenched\cite{Coleman}. However, if the Kondo lattice system is disordered, we expect a remnant magnetic moment to exist in the system, and this downturn due to coherence can be averted. We expect that this remnant magnetic moment acts to produce effective Kondo scattering, and the logarithmic increase of resistivity can still remain. Indeed, there are examples of heavy fermion systems that show suppression of the downturn by introducing even a small doping amount\cite{Scoboria,Amakai}, i.e. small disorder. 

The downturn of $\mu(B)$ at higher magnetic fields is not a feature of Kondo scattering, but is qualitatively consistent with short-range disorder scattering mechanisms. It has long been known\cite{Ando} that in the high carrier density limit (in which the SmB$_6$ surface states live), the mobility is partially determined by short range disorder scattering mechanisms (e.g., surface roughness scattering) and scales inversely with the carrier density, i.e., $\mu\propto n^{-\alpha}$, where $\alpha$ is determined by the particular scattering mechanism(s). This behavior has been observed in several semiconductor heterostructures\cite{Manfra}. In SmB$_6$, as $n(B)$ increases with increasing $B$, the short-range scattering time (and thus $\mu$) decreases with $B$. Typical values for $\alpha$ between $\frac{1}{2}$ and 2 are consistent with our data. However, a precise determination of $\alpha$ from the data is problematic, because $n(B)$ only varies by 15\% over the fields measured, and the dynamic range for determining a power-law relation is too small. This further complicates the determination of ${T_\text{K}}$, since the contribution to the mobility from the short-range scattering can compensate the contribution from the Kondo effect. (For example, for an $\alpha$ of $\frac{1}{2}$ , a ${T_\text{K}}$ of 30 K would give a better fit in Figure~\ref{fig:Kondo}.) However, this effect, together with the Kondo scattering, gives a picture that is qualitatively consistent with the $\mu(B)$ we extract from our analysis, where the low-field negative MR is due mostly to Kondo scattering, and the high-field negative MR is due mainly to an enhancement in $n(B)$, which then causes a much weaker diminution of $\mu(B)$ via an increase in short-range scattering.

\subsection{\label{sec:WAL}Weak Anti-localization}

Consistent with our results for the temperature dependence and magnetic field dependence of the resistivity, a possible reason for WAL to be absent or too small to measure is because of the existence of magnetic (Kondo) impurity scattering. Magnetic impurity scattering plays a role in the quantum correction of conductivity, since it alters the dephasing of electrons. Here, we expect two possible effects that can reduce the magnitude of the dips. For non-TI 2DEGs, it is well known that introducing a small number of magnetic impurities can even switch the signs of the dips of the quantum correction to conductivity\cite{Bergmann1984}. For a topological insulator surface, there is an additional effect that results in a smaller feature size. By introducing magnetic impurities, the energy band gap at the Dirac point opens, and this band gap opening induces a crossover from weak anti-localization to weak localization (WL)\cite{HZLu}. These considerations, which are entirely expected in a system with magnetic impurities, fall outside the scope of the usual Hikami-Larkin-Nagoaka formulation\cite{HikamiLarkinNagoaka} used to analyze WL and WAL.

Finally, we mention that the magnetic field range appropriate for a quantum correction to conductivity on the SmB$_6$ surface may not be the typical range for such corrections. Ordinarily for disordered thin metals and other known TI surfaces\cite{Taskin,Roy}, the magnetic field range of interest for WAL or WL is $0.1\text{ -- }1\text{ T}$. Theoretically, this range can be estimated by the characteristic field $B_{\phi}=(h/e)/8\pi l_{\phi}^2$ without magnetic impurity scattering. Since even a small amount of magnetic impurities can lower the phase coherence length by orders of magnitude\cite{CoherenceMagImpur1,CoherenceMagImpur2}, the characteristic field can be much larger than 1 T. In future studies, a wider range of magnetic field may need to be considered for WAL and WL. 

\subsection{\label{sec:DynMagFeat}Origin of the Dynamical Magnetotransport Feature}

We now address the physical origin of the hysteretic features seen at low field values. Because SmB$_6$ exhibits no magnetic ordering at low temperatures, a magnetic hysteresis is likely to be extrinsic to the material, arising from the surface oxide (which may vary significantly across our samples) or some other material used to mount the samples in our cryosystems and proximal to our samples. All the samples we measured exhibited the very low field ($\sim0.05$ T) peaks in multiple cryostats, but only some exhibited the higher-field $(0.5\text{ -- }2\text{ T})$ peaks.

We note that trivial heating from the sample or from an external source (e.g. the magnet power supply) cannot explain this behavior. First, Joule heating of the sample cannot be the case. By changing the current through the sample by an order of magnitude, we did not observe a change in the hysteretic behavior. Also, Joule heating of the sample is orders of magnitude smaller than the cooling power of our cryogenic system. Second, inductive heating by eddy currents cannot be the case. Inductive heating depends on the magnetic field sweep rate, but is independent of the sweep direction. Since inductive heating is constant throughout a fixed-sweep rate, if inductive heating causes the resistivity change of the sample, this change can only be monotonic and non-reproducible over several sweep cycles. However, our data have two non-monotonic dips which are reproducible at a constant sweep rate and temperature. Also, comparing to the cooling power at $0.3\text{ K}$, the magnitude of inductive heating is orders of magnitude smaller. We also observe that the temperature fluctuations recorded in our thermometer are not large enough to indicate a global temperature change in the system. Finally, if a single polar power supply is used for the superconducting magnet, it can cause a dip in resistivity as it switches the circuit at zero magnetic field. For this reason, we used a bipolar magnetic power supply for which the switching event ($B\neq0\text{ T}$) was identified, and we confirmed that the dips are independent from this event. 

One possible extrinsic source of this behavior could be the magnetocaloric effect. In this scenario, the increase in magnetic field coerces the magnetic moments in a magnetic material to align with the field, which reduces the magnetic entropy of the system. We expect this process to be adiabatic in our experiment, leading to an increase in temperature. When the magnetic field is then reduced toward zero, the magnetic entropy can increase, leading to a decrease in the sample temperature. Such variations in the sample temperature would change the resistance according to Figure~\ref{fig:Temp}. Indeed, for most of our samples, the decrease in resistance as the field is increased would be consistent with a temporary increase in temperature. As the magnetization becomes saturated at higher fields, this warming effect would gradually disappear, allowing the cryosystem to cool the sample over a timescale of several seconds, consistent with our observations, and allowing the resistance to return to its original value. If the source of the magnetization is located very near the sample, it may be sufficiently thermally isolated from the thermometer and cooling power of the system to influence the sample temperature without influencing the thermometer.

The magnetic material that would be responsible for this effect is entirely unclear. SmB$_6$ itself does not exhibit magnetic ordering at low temperatures. The electronic leads to our samples include a number of possible superconducting materials, but these typically exhibit the inverse magnetocaloric effect, which has the wrong sign for our observations, and typically have critical temperatures well within or below the temperature ranges in which we observe this effect. We also occasionally observe spasmodic, non-reproducible spikes in the data and an offset in the resistance that depends on sweep direction to fields as high as 8 T, well above the critical field for most superconductors. However, it is not likely that the magnetocaloric effect is responsible for these sporadic features.

Another possibility is that this hysteretic magnetotransport behavior with long time scales is related to the disordered Kondo lattice system of Sm$_2$O$_3$ on the SmB$_6$ surface. In addition to the picture introduced in Section~\ref{sec:Kondo}, we posit that the disordered Kondo lattice is also a glassy system\cite{Theumann, Magalhaes}. Previously studied disordered Kondo lattice models that are glassy assume that RKKY interaction strengths are randomly distributed. Here we also point out the possibility that random superexchange interactions due to the varying angle of the Sm-oxygen-Sm bonds may play a role similar to that of random RKKY interactions. Our measured resistivity response to the low magnetic field may be a manifestation of the magnetization property of a glassy system. In a spin glass system, the relaxation time of the magnetization can be extremely long\cite{BinderSpinGlassRMP}. Therefore, when an external magnetic field is applied, the magnetization depends on the magnetic field sweep rate, even at very slow rates. In addition, the total magnetization of a spin glass system exhibits a hysteresis loop, so the area of the hysteresis loop depends on the magnetic field sweep rate. Theoretically, the hysteresis area becomes larger at faster sweep rates, and at lower temperatures\cite{Sariyer}. In a scenario where the resistivity decreases when the magnetization decreases, the magnetic field history, sweeping direction, sweep rate, and temperature dependence of our data are all consistent with the magnetization of the glassy features explained above.

This glassy magnetic ordering scenario along with the absence of WAL is inconsistent with previous reports\cite{SThomasWAL, NakajimaJP}. Both previous studies report observing WAL, and Nakajima \textit{et al}. additionally reports on sweep-rate-independent hysteresis as evidence of chiral edge channels from ferromagnetic domains\cite{NakajimaJP}. One possible explanation for this difference is that the surface magnetic ordering of our samples is quite different from that of the samples in those experiments, due to variations in the disorder of the native oxide after different sample preparation procedures, such as polishing and lithography. According to the disordered Kondo lattice model, the magnetic phase can change between spin glass and ferromagnetic ordering, depending on the degree of disorder\cite{Magalhaes}. Further systematic studies of surface preparation are needed to reconcile these differing findings. In addition, transport measurements performed in high-vacuum on cleaved surfaces, on which there is presumably no oxide layer, would also be extremely powerful for the full characterization of the surface states.

\section{\label{sec:Conclusion}Conclusion}

We have performed transport measurements of individual crystallographic surfaces of SmB$_6$ using Corbino disk structures. Both (001) and (011) surfaces display strong negative MR. The (011) surface exhibits a carrier density and mobility at values which are significantly lower than previously reported from transport methods, but which are more consistent with ARPES data. For both (001) and (011) surfaces, the temperature dependence suggests Kondo scattering from magnetic surface impurities. Fits of the angular dependence of our data suggest that the negative MR is primarily due to an increase in carrier density, especially at high field, but with some additional contribution from the suppression of Kondo scattering.

All of our samples revealed a dip of resistivity which depended on the magnetic field sweep rate. Although these features become smaller in magnitude at slower sweep rates, the magnitude is still clearly visible at our slowest measurements. These features are most likely due to an extrinsic magnetic effect such as the magnetocaloric effect or magnetic impurity scattering due to the presence of the naturally formed samarium oxide (Sm$_{2}$O$_{3}$) layer, which might exhibit a glassy magnetic ordering. In either case, the behavior of the dip is inconsistent with WAL, and to the extent permitted by the dip at the slowest sweep rates, we do not observe WAL. This lack of WAL could also be attributed to the effect of the magnetic surface impurities.

A topological insulator with no bulk contribution can potentially be an ideal building block for realizing Majorana Fermions and spintronics devices\cite{Wilczek, Fu, Yokoyama}. If the Sm$_2$O$_3$ is the leading magnetic impurity on the surface, the impurities, in principle, can be avoided using oxygen-free fabrication conditions . In this case, the surface of SmB$_{6}$ may be a strong candidate for this building block. Growing a heterostructure or a cap layer on top of the SmB$_{6}$ surface may be a possible solution for preventing the native samarium oxide formation on the SmB$_{6}$ surface.

\begin{acknowledgments}
We wish to acknowledge Kyunghoon Lee for his assistance with wirebonding the Corbino contacts, Juniar Lucien for polishing crystal surfaces, and Jan Jaroszynski for discussion of the magnetocaloric effect. This work was supported by the National Science Foundation grants \# ECCS-1307744, DMR-1006500, DMR-1441965, and DMR-0801253, the Department of Energy award DE-SC0008110, the Scientific and Technological Research Council of Turkey (TUBITAK), the China Scholarship Council, and the National Basic Research Program of China (973 Program, Grant No. 2012CB922002). Device fabrication was performed in part at the Lurie Nanofabrication Facility, a member of the National Nanotechnology Infrastructure Network, which is supported by the National Science Foundation. The high-field experiments were performed at the National High Magnetic Field Laboratory, which is supported by NSF Cooperative Agreement No. DMR-084173, by the State of Florida, and by the DOE.
\end{acknowledgments}

\appendix

\section{\label{apx:001Cond}Analysis of the (001) Surface Conductivity}

\begin{figure}
\includegraphics{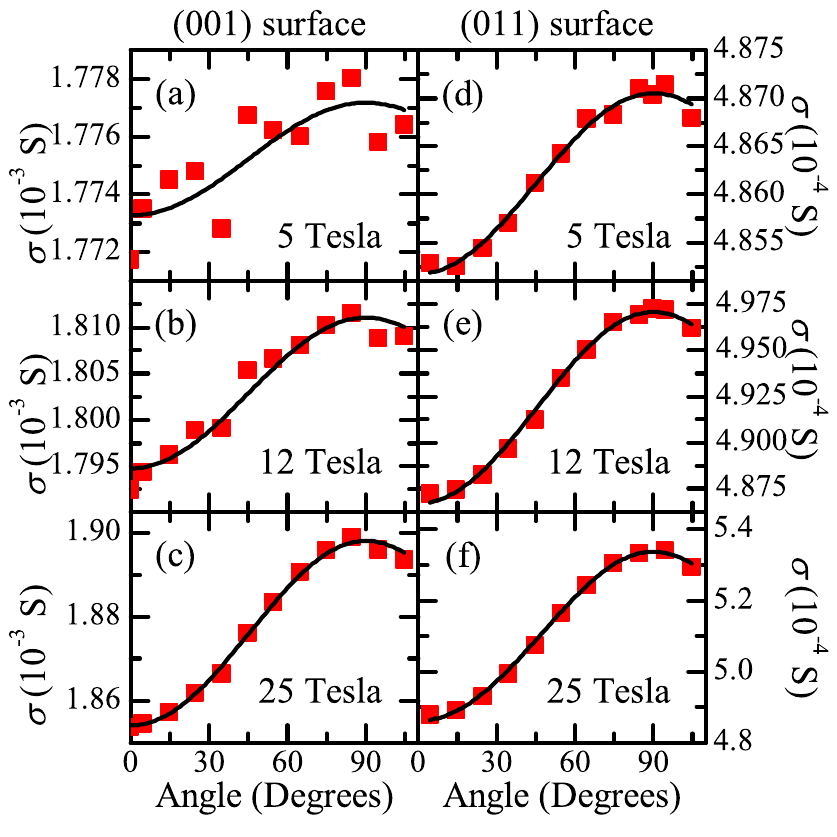}
\caption{(Color online) Cosine fits (solid curves) of angle-dependent magnetoresistance data (red squares) at 5, 12, and 25 T magnetic field, respectively, for the (001) surface (a)--(c) and the (011) surface (d)--(f).}
\label{fig:cosine}
\end{figure}

As discussed in the main body of the article, the angle-dependent magnetoresistance (MR) of the (001) surface states at 0.3 K and below $\sim20\text{ T}$ do not exhibit the $\cos^2\theta$ behavior expected from the Corbino geometry. Figure~\ref{fig:cosine}~(a)--(c) show the (001) MR as a function of $\theta$ for constant magnetic field at 5, 12, and 25 T, respectively. At 5 T, the amplitude of the cosine fit is small compared to the variance in the data; thus, the uncertainty in our fitting parameters is dominated by noise, a problem which becomes worse as $B^2\to0$. At 25 T, the $\cos^2\theta$ fit is quite good (the residuals are quite small), and the uncertainty in the fitting parameters is very small. However, at 12 T, the data deviates somewhat from the $\cos^2\theta$ fit, suggesting that some other angle-dependent mechanism is influencing the conductivity. We see this behavior from $\sim7\text{ T}$ to $\sim25\text{ T}$ in the (001) surface only. The $\cos^2\theta$ fits are quite good above 25 T for the (001) surface, as well as all field values for the (011) surface, except near where $B^2\to0$, as expected. Figure~\ref{fig:cosine}~(d)--(f) show the $\cos^2\theta$ fits for the (011) surface at the same $B$-values for comparison.

\begin{figure}
\includegraphics{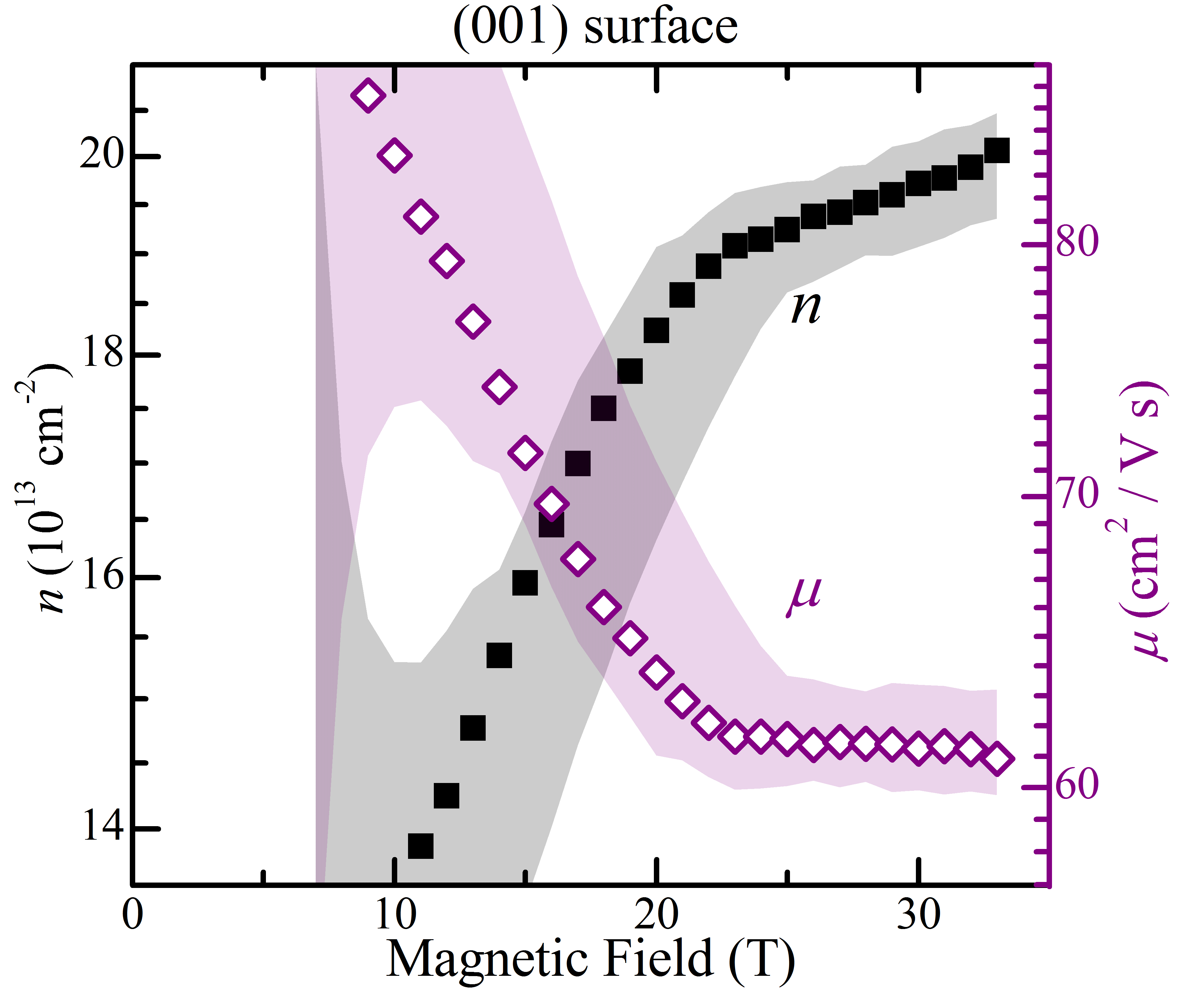}
\caption{(Color online) (001) surface carrier density (filled squares) and mobility (open diamonds) obtained from angle-dependent fits of the data. Shaded areas represent uncertainty in the parameters of the angle-dependent fits. The vertical log scale allows direct comparison of the relative magnitudes of changes in $n$ and $\mu$, and is proportional to Figure~2 of the article. We warn the reader that these values are not reliable, as discussed in the text.}
\label{fig:AngleFits001}
\end{figure}

We prepared a summary of our fits for $n(B)$ and $\mu(B)$ for the (001) surface in Figure~\ref{fig:AngleFits001}. The distinct regimes of behavior discussed are evident in the plot. Above $\sim25\text{ T}$, $\mu(B)$ is relatively constant, and has a fit uncertainty (shaded region in Figure~\ref{fig:AngleFits001}) comparable to that of the (011) surface. In this regime, $n(B)$ also increases, similarly to the (011) surface carrier density. Below $\sim25\text{ T}$, the calculated values for $n(B)$ and $\mu(B)$ change dramatically, and the residuals of the fits (and the corresponding fit uncertainties) become quite large. Below $\sim7\text{ T}$, the fits become completely unreliable, as indicated by the diverging uncertainties in $n(B)$ and $\mu(B)$.

It is likely that the (001) surface has at least two carrier types, but this possibility does not really answer why the middle regime differs a bit from both the $\cos^2\theta$ behavior and the $B^2$ behavior while the high-field regime follows both behaviors quite well. If the effect responsible for this difference is limited to below $\sim25\text{ T}$, then the values and trends obtained for $n(B)$ and $\mu(B)$ above $25\text{ T}$ may still be useful. In any case, there is no way to distinguish single carrier conduction from multi-carrier conduction with carriers of similar mobilities, either on the (001) surface or the (011) surface.

\section{\label{apx:002Kondo}Kondo scattering as the origin of the magnetoresistance}

\begin{figure}
\includegraphics{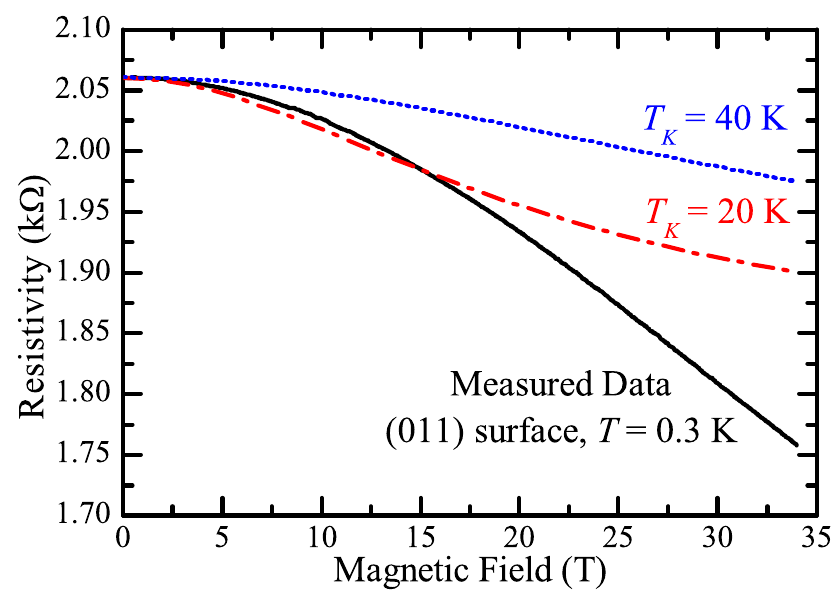}
\caption{(Color online) Simulated magnetoresistance of the (011) surface due to Kondo scattering alone, based on the values obtained from the logarithmic temperature dependence. Simulations for Kondo temperatures of 20~K (red dash-dot curve) and 40~K (blue dotted curve) are shown alongside the actual magnetoresistance obtained from measurements (solid black curve).}
\label{fig:KondoFail}
\end{figure}

The logarithmic increase in the surface resistivity as temperature drops is an indication for Kondo scattering due to magnetic impurities near the surface of the crystal. This is the most likely origin of the logarithmic increase, since the logarithmic coefficient is not near $e^2/h$, as would be expected for quantum interference effects. We initially investigated this as a candidate for negative MR, since a magnetic field suppresses the formation of Kondo singlets, thereby reducing the scattering rate due to magnetic impurities. However, the maximum reduction of the scattering rate corresponds to zero contribution from the Kondo mechanism---turning Kondo scattering off. The contribution of Kondo scattering to the increase in resistivity at $0.3\text{ K}$ can be estimated from the difference between the actual resistivity and the background resistivity. If the magnetic field ``turns off'' the Kondo scattering, the resistivity should drop to the background level. However, we observe that the magnetic field reduces the resistivity well beyond this limit. This is easily seen in Figure~\ref{fig:KondoFail}, which shows our MR data alongside simulations of the Kondo scattering predicted by the temperature dependence. The Kondo temperature $T_\text{K}$ is an adjustable parameter, so we give plots for two values which respectively under- and over-estimate the low-field MR. However, neither value correctly captures the high-field MR observed---the magnitude of the effect at high field is smaller than the magnitude of the observed MR. Thus, the Kondo scattering mechanism alone cannot explain the magnitude of the MR we observe.

\providecommand{\noopsort}[1]{}\providecommand{\singleletter}[1]{#1}%

\end{document}